\documentclass[11pt]{article}

\usepackage{arxiv}

\usepackage[utf8]{inputenc} 
\usepackage[T1]{fontenc}    
\usepackage{hyperref}       
\usepackage{url}            
\usepackage{booktabs}       
\usepackage{amsfonts}       
\usepackage{nicefrac}       
\usepackage{microtype}      
\usepackage{lipsum}
\usepackage{graphicx}
\usepackage{physics}
\usepackage{subcaption}
\usepackage{authblk}

\title{OpticalGAN : Generative Adversarial Networks for Continuous Variable Quantum Computation}

\author[*]{Nilay Shrivastava}
\author[*]{Nikaash Puri}
\author[*]{Piyush Gupta}
\author[*]{Balaji Krishnamurthy}
\author[*]{Sukriti Verma}

\affil[*]{Media and Data Science Research Lab, Adobe}

\begin{document}
\maketitle

\begin{abstract}
We present OpticalGAN, an extension of quantum generative adversarial networks for continuous-variable  quantum computation. OpticalGAN consists of photonic variational circuits comprising of optical Gaussian and Kerr gates. Photonic quantum computation is a realization of continuous variable quantum computing which involves encoding and processing information in the continuous quadrature amplitudes of quantized electromagnetic field such as light. Information processing in photonic quantum computers is performed using optical gates on squeezed light. Both the generator and discriminator of OpticalGAN are short depth variational circuits composed of gaussian and non-gaussian gates. We demonstrate our approach by using OpticalGAN to generate energy eigenstates and coherent states. All of our code is available at https://github.com/abcd1729/opticalgan.
\end{abstract}

\section{Introduction} \label{sec:introduction}
The Quantum Generative adversarial networks (GAN) for the qubit model of quantum computation is described in \cite{qugan1} where the generator and discriminator are modelled as quantum circuits composed of unitary gates. An important challenge for Quantum GANs is to effectively implement affine and non-linear transforms. Attempts have been made to induce non-linearity indirectly \cite{qubitdisadv3, qubitdisadv4} but the resulting circuit has significant failure probability \cite{circuitclassifier, cvnn}. Further, Qubit based quantum computers are only partially suitable for continuous valued problems since the measurement output of the circuits is discrete \cite{cvnn, qubitdisadv1, qubitdisadv2}. Continuous variable (CV) quantum computation \cite{cvbegin1, cvbegin2} is an alternative to the qubit model. The state of the smallest unit of CV computation, \textit{qumode}, is part of an infinite dimensional Hilbert space. Information is encoded in quantum states of fields and the system is suitable for implementation on photonic hardware \cite{cvintro1, cvintro2, cvintro3}. 
It is shown in \cite{cvbegin2} that the qubit model can be implemented using such a photonic system without any loss of universality. Moreover, when the continuous quadrature of canonical position ($\hat{x}$) and canonical momentum ($\hat{p}$) operators are used for information processing, the CV model lends itself for constructing neural networks \cite{cvnn}.

We present OpticalGAN, an extension of Quantum GAN for continuous variable quantum computation. We use quantum variational circuits to simulate neural networks. Affine transformations are implemented using Gaussian operations on the initial vacuum state and non-linearity is implemented using the Kerr and Cubic phase gates \cite{quantumoptics, stateprep}. The generator and discriminator are implemented using these variational circuits and trained in an adversarial setting as described in \cite{qugan1}. During training the parameters of the variational circuit are optimized using backpropagation to minimize the adversarial loss \cite{gan}. The contributions of our work are:
\begin{itemize}
  \item We propose \textbf{OpticalGAN}, a generative adversarial network formulated for continuous variable (CV) quantum computation. The quantum circuits of OpticalGAN simulate generator and discriminator networks in CV settings using optical gates. (Section \ref{sec:opticalgan}).
  
  \item We show the efficacy of OpticalGAN by training it to \textbf{generate fock and coherent states } (Section \ref{sec:expresults}). Our experiments show that during adversarial training, the cross-entropy loss between the generated and target distribution tends to zero. 
\end{itemize}

\section{OpticalGAN} \label{sec:opticalgan}

\begin{figure}
\centering
\begin{subfigure}{.5\textwidth}
  \centering
  \includegraphics[width=1\linewidth]{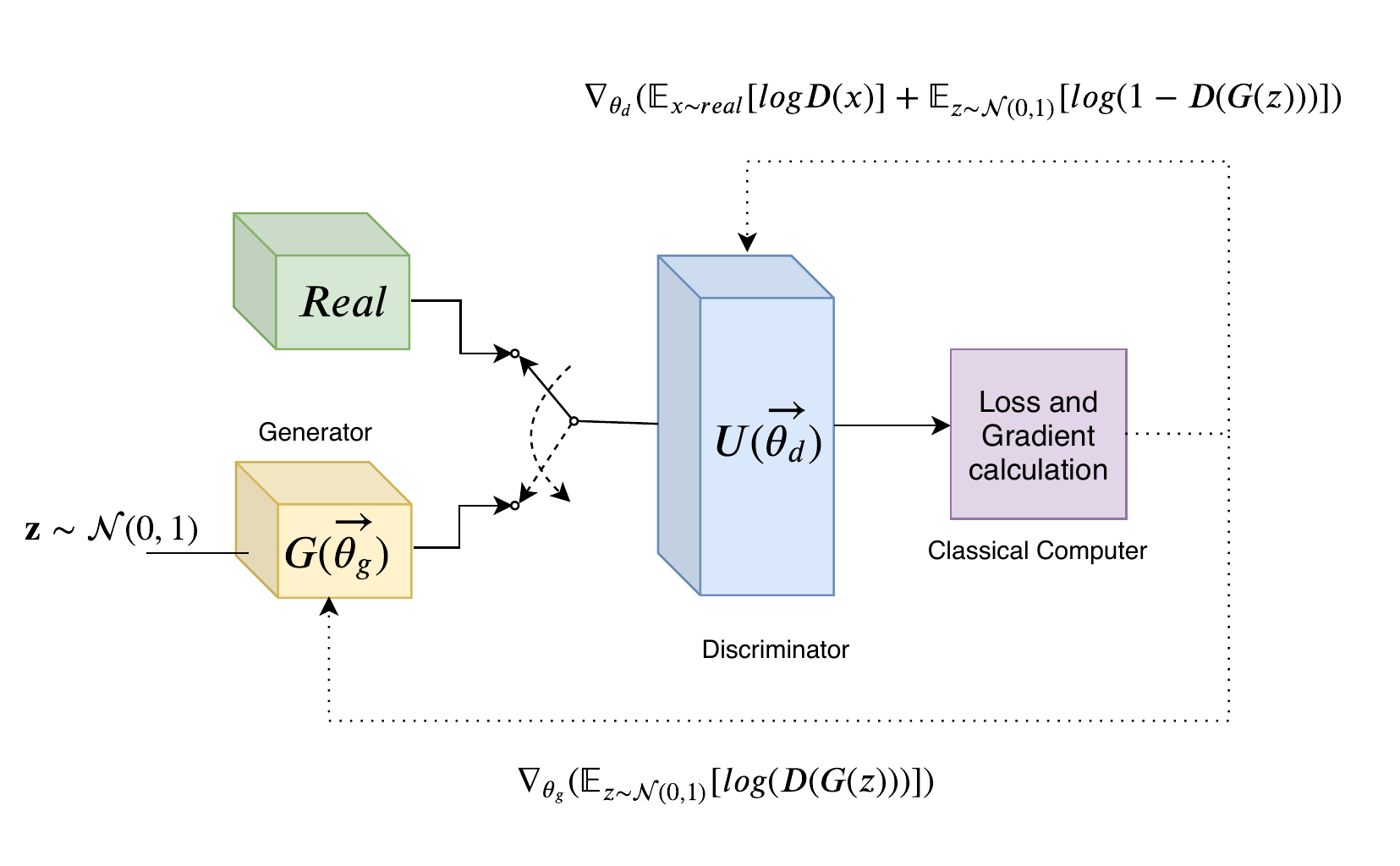}
  \caption{Architecture of OpticalGAN}
    \label{fig:opticalgan}
\end{subfigure}%
\begin{subfigure}{.5\textwidth}
  \centering
  \includegraphics[width=0.5\linewidth]{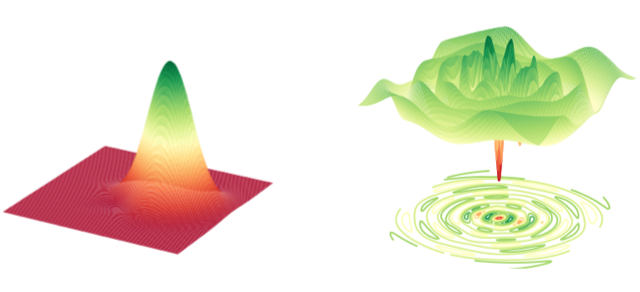}
  \caption{Generated coherent state with $\alpha=1$ and Generated Fock state $\ket{9}$}
  \label{fig:coherentgen}
\end{subfigure}

\caption{Architecture and Results}
\label{fig:test}
\end{figure}

The architecture of OpticalGAN is shown in Fig. \ref{fig:opticalgan}. There are four components, a real data source, generator and discriminator quantum circuits and a classical computer. The data source generates quantum states that are representative of the real data distribution. The mapping from classical data to quantum states can be achieved as in \cite{cvnn}. We assume that the quantum state generated can be represented using a single qumode for the sake of simplicity. The following sections first discuss the basic building block of OpticalGAN, the quantum neural network layer \cite{cvnn}. We then describe the generator, discriminator variational circuits and finally the cost function and training paradigm. 

In order to implement a quantum neural network layer $\mathcal{L}$, we use \cite{cvnn}. The essential idea is that by using interferometers, squeeze gates (denoted by $S$), displacement gates ($D$) and a non-gaussian Kerr gate ($K$), one can create quantum neural networks. We use the layer circuit described in \cite{cvnn} as a component in the generator and discriminator variational circuits.

\subsection{Generator and Discriminator variational circuits} \label{sec:generator}
\begin{figure}
\centering
\begin{subfigure}{.5\textwidth}
  \centering
  \includegraphics[width=1\linewidth]{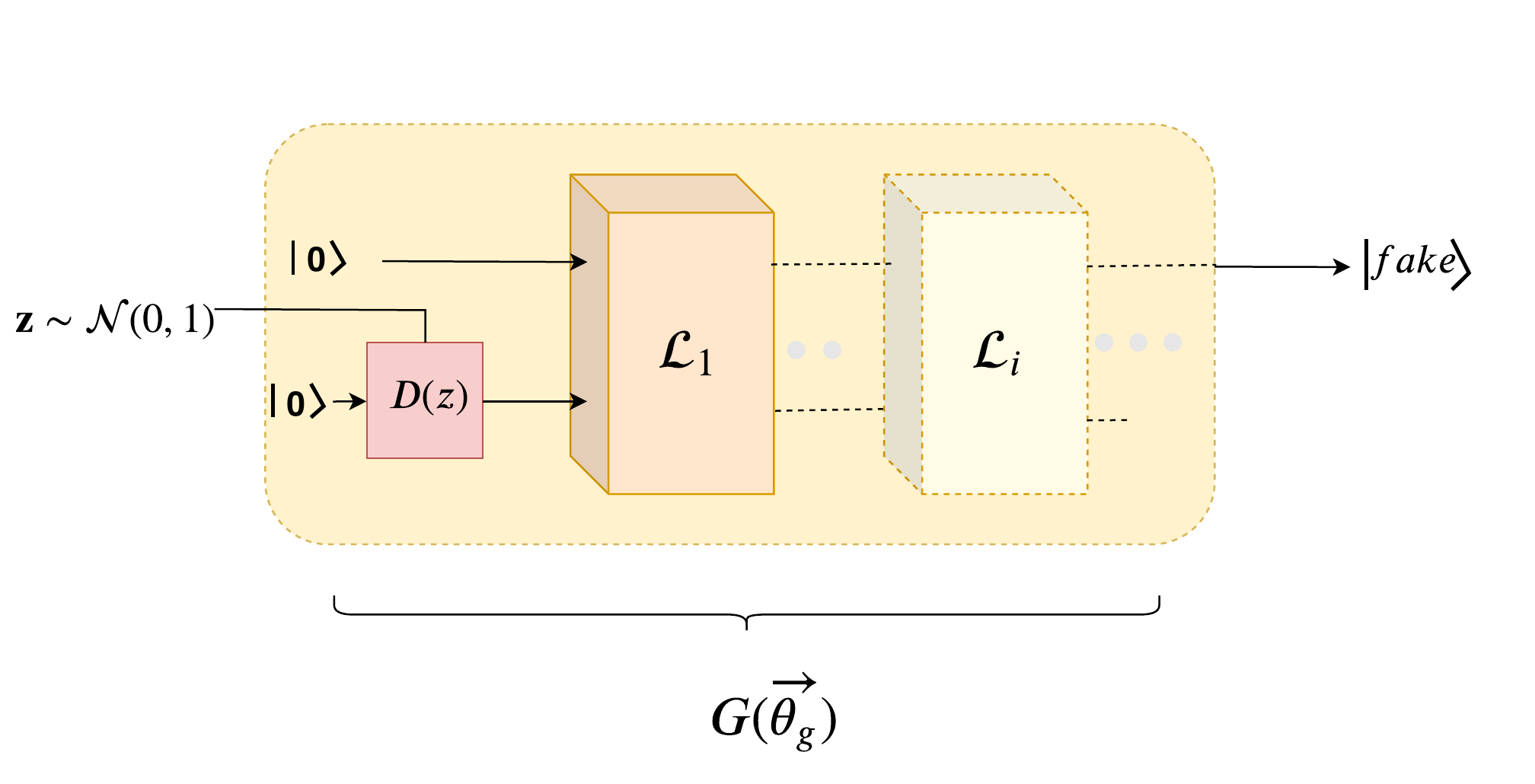}
  \caption{Design of the Generator circuit}
  \label{fig:generator}
\end{subfigure}%
\begin{subfigure}{.5\textwidth}
  \centering
  \includegraphics[width=1\linewidth]{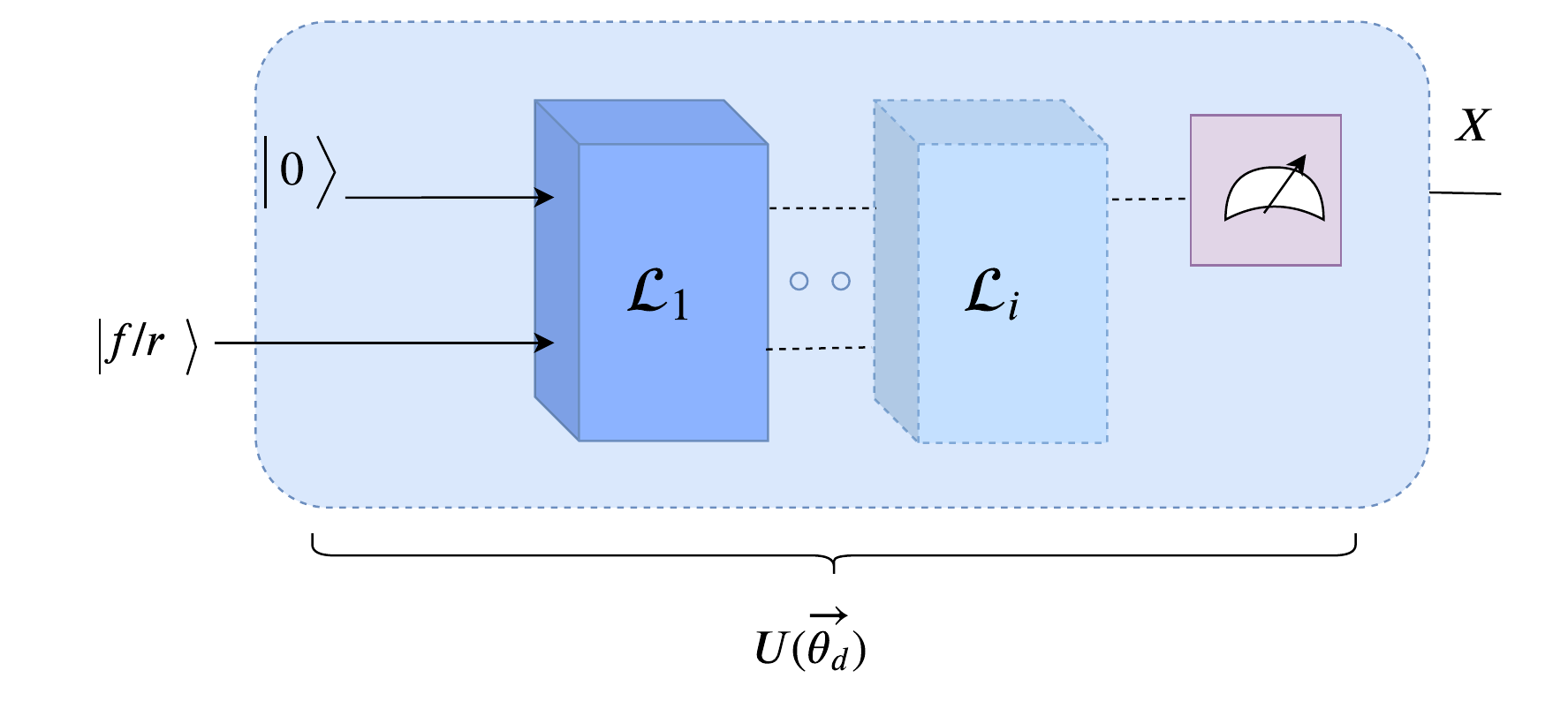}
  \caption{Design of the Discriminator circuit}
  \label{fig:discriminator}
\end{subfigure}
\caption{Design of the Generator and Discriminator circuits}
\label{fig:test}
\end{figure}

The generator is a variational circuit consisting of two qumodes and the output is obtained from one of the two qumodes, as shown in Fig. \ref{fig:generator}. It takes as input a gaussian random variable $z$ as input and produces a quantum state. $z$ is used as a parameter for the displacement gate $D$. The gate is applied on one of the two qumodes that are instantiated in the vacuum state $\ket{0}$. This is followed by variational sub-circuits denoted by $\mathcal{L}$ \cite{cvnn}. Since the $\mathcal{L}$'s can be stacked, the variational circuit can be thought of as a quantum neural network with several layers. $\vec{\theta_g}$ denotes the parameters of the generator circuit. $\vec{\theta_g}$ is tuned during training so that the generator produces an output that matches the distribution of the real data.

The discriminator circuit is similar to the generator. The key difference is that it takes as input the state produced by either the real source or the generator and tries to distinguish them. It uses quantum neural network layers \cite{cvnn} as variational sub-circuits. $\vec{\theta_d}$ denotes the parameters. At the end of the circuit we perform a homodyne measurement to calculate the expected value of the $\hat{x}$ quadrature. Given the expected value ($X$), the probability that the input state is real is:
\begin{equation} \label{eq:prob}
    D(input) = P(input \sim real; \vec{\theta_d}) = \frac{1}{1 + e^{-X}}
\end{equation}

The loss function is computed using a classical computer. The output of the discriminator, i.e. the expectation value $X$ is used for computing the GAN loss \cite{gan} 
\begin{equation}
    L_{D} = \mathbb{E}_{x \sim real}[log D(x)] + \mathbb{E}_{z \sim \mathcal{N}(0,1)[log(1-D(G(z)))]} \quad and \quad
    L_{G} = \mathbb{E}_{z \sim \mathcal{N}(0,1)}[log(D(G(z)))]
\end{equation}
Here $x$ is a data point from the real distribution, $G(z)$ is the output of the generator. $D(x)$ is the probability $P(x \sim real; \vec{\theta_d})$ given in Eq. \ref{eq:prob}. We follow the training protocol proposed in \cite{qugan1}. Gradient calculation is done using automatic differentiation \cite{paramshift}. We use this to perform quantum gradient descent to find the optimal values of $\vec{\theta_g}$ and $\vec{\theta_d}$.

\section{Experiments and Results} \label{sec:expresults}
The experiments have been performed using Pennylane and Strawberryfields \cite{pennylane, strawberryfields}. We use OpticalGAN to generate coherent states which are at unit distance from vacuum state in phase space and to generate energy eigentates (fock basis). Due to limited space, Fig. \ref{fig:coherentgen} shows both the generated coherent state for $\alpha=1$ and the generated fock state $\ket{9}$. For more details and results, refer to the code available at https://github.com/abcd1729/opticalgan. We are currently working on generating images using OpticalGAN. 


\clearpage

\bibliographystyle{unsrt}
\bibliography{opgan}

\begin{thebibliography}{10}

\bibitem{qugan1}
Pierre-Luc Dallaire-Demers and Nathan Killoran.
\newblock Quantum generative adversarial networks.
\newblock {\em Physical Review A}, 98(1):012324, 2018.

\bibitem{qubitdisadv3}
Eric Torrontegui and Juan~Jos{\'e} Garcia-Ripoll.
\newblock Universal quantum perceptron as efficient unitary approximators.
\newblock {\em arXiv preprint arXiv:1801.00934}, 2018.

\bibitem{qubitdisadv4}
Yudong Cao, Gian~Giacomo Guerreschi, and Al{\'a}n Aspuru-Guzik.
\newblock Quantum neuron: an elementary building block for machine learning on
  quantum computers.
\newblock {\em arXiv preprint arXiv:1711.11240}, 2017.

\bibitem{circuitclassifier}
Maria Schuld, Alex Bocharov, Krysta Svore, and Nathan Wiebe.
\newblock Circuit-centric quantum classifiers.
\newblock {\em arXiv preprint arXiv:1804.00633}, 2018.

\bibitem{cvnn}
Nathan Killoran, Thomas~R Bromley, Juan~Miguel Arrazola, Maria Schuld,
  Nicol{\'a}s Quesada, and Seth Lloyd.
\newblock Continuous-variable quantum neural networks.
\newblock {\em arXiv preprint arXiv:1806.06871}, 2018.

\bibitem{qubitdisadv1}
Marcello Benedetti, John Realpe-G{\'o}mez, and Alejandro Perdomo-Ortiz.
\newblock Quantum-assisted helmholtz machines: A quantum--classical deep
  learning framework for industrial datasets in near-term devices.
\newblock {\em Quantum Science and Technology}, 3(3):034007, 2018.

\bibitem{qubitdisadv2}
Alejandro Perdomo-Ortiz, Marcello Benedetti, John Realpe-G{\'o}mez, and Rupak
  Biswas.
\newblock Opportunities and challenges for quantum-assisted machine learning in
  near-term quantum computers.
\newblock {\em Quantum Science and Technology}, 3(3):030502, 2018.

\bibitem{cvbegin1}
Emanuel Knill, Raymond Laflamme, and Gerald~J Milburn.
\newblock A scheme for efficient quantum computation with linear optics.
\newblock {\em nature}, 409(6816):46, 2001.

\bibitem{cvbegin2}
Daniel Gottesman, Alexei Kitaev, and John Preskill.
\newblock Encoding a qubit in an oscillator.
\newblock {\em Physical Review A}, 64(1):012310, 2001.

\bibitem{cvintro1}
Seth Lloyd and Samuel~L Braunstein.
\newblock Quantum computation over continuous variables.
\newblock In {\em Quantum Information with Continuous Variables}, pages 9--17.
  Springer, 1999.

\bibitem{cvintro2}
Samuel~L Braunstein and Peter Van~Loock.
\newblock Quantum information with continuous variables.
\newblock {\em Reviews of Modern Physics}, 77(2):513, 2005.

\bibitem{cvintro3}
Samuel~L Braunstein and Peter Van~Loock.
\newblock Quantum information with continuous variables.
\newblock {\em Reviews of Modern Physics}, 77(2):513, 2005.

\bibitem{quantumoptics}
Peter Lambropoulos and David Petrosyan.
\newblock {\em Fundamentals of quantum optics and quantum information},
  volume~23.
\newblock Springer, 2007.

\bibitem{stateprep}
Juan~Miguel Arrazola, Thomas~R Bromley, Josh Izaac, Casey~R Myers, Kamil
  Br{\'a}dler, and Nathan Killoran.
\newblock Machine learning method for state preparation and gate synthesis on
  photonic quantum computers.
\newblock {\em Quantum Science and Technology}, 2018.

\bibitem{gan}
Ian Goodfellow, Jean Pouget-Abadie, Mehdi Mirza, Bing Xu, David Warde-Farley,
  Sherjil Ozair, Aaron Courville, and Yoshua Bengio.
\newblock Generative adversarial nets.
\newblock In {\em Advances in neural information processing systems}, pages
  2672--2680, 2014.

\bibitem{paramshift}
Gavin~E Crooks.
\newblock Gradients of parameterized quantum gates using the parameter-shift
  rule and gate decomposition.
\newblock {\em arXiv preprint arXiv:1905.13311}, 2019.

\bibitem{pennylane}
Ville Bergholm, Josh Izaac, Maria Schuld, Christian Gogolin, and Nathan
  Killoran.
\newblock Pennylane: Automatic differentiation of hybrid quantum-classical
  computations.
\newblock {\em arXiv preprint arXiv:1811.04968}, 2018.

\bibitem{strawberryfields}
Nathan Killoran, Josh Izaac, Nicol{\'a}s Quesada, Ville Bergholm, Matthew Amy,
  and Christian Weedbrook.
\newblock Strawberry fields: A software platform for photonic quantum
  computing.
\newblock {\em Quantum}, 3:129, 2019.

\end{thebibliography}

\end{document}